\documentstyle[aps,preprint,tighten]{revtex}

\begin{document}


\preprint{\vbox{Submitted to Physics Letters B
                \hfill DOE/ER/40762--037\\
                \null\hfill UMPP \#94--136}}

\title{QCD sum rules vs.\ chiral perturbation theory}
\author{David K. Griegel and Thomas D. Cohen}
\address{Department of Physics and Center for Theoretical Physics\\
University of Maryland, College Park, Maryland 20742}
\date{May 1994}
\maketitle
\begin{abstract}
QCD sum rules are useful tools for studying the spectral properties of
hadrons;
however, assumptions underlying standard sum-rule analyses can lead to
inconsistencies with known results of chiral perturbation theory.
This possibility is demonstrated with QCD sum-rule extractions of the
nucleon mass and $\sigma$ term.
In both cases, inconsistent chiral behavior leads to an uncertainty
in the sum-rule predictions of $\sim 100\,\text{MeV}$.
\end{abstract}
\pacs{PACS numbers: ???}


QCD sum rules are based on the analytic properties of current-current
correlators and asymptotic freedom \cite{shifman,reinders}.
Hadronic spectral properties are related via dispersion relations to
the current-current correlator evaluated in terms of fundamental
quark and gluon fields.
This evaluation uses an operator product expansion (OPE), which, in
turn, requires knowledge of various vacuum matrix elements---the
condensates.
In practice, applications of QCD sum rules require simple parameterizations
of spectral functions and a truncated OPE.
These simplifications necessarily lead to uncertainities in the extracted
spectral parameters.
Moreover, it is difficult to assess the magnitude of these
uncertainties reliably.
Accordingly, it is useful to test QCD sum-rule results against reliable
physical and mathematical constraints.
In this Letter, we will use well-known properties of hadronic spectra in
the chiral limit to test the dependability of QCD sum-rule extractions
of the nucleon mass and $\sigma$ term.

It is generally believed that chiral perturbation theory accurately
describes all low-energy observables of QCD in the limit of light
current quark masses \cite{gasser1,gasser2}.
In the present context, however, we will only be interested in the
leading nonanalytic behavior of observables as a function of $m_q$, the
average of the up and down current quark masses.
It should be noted that this nonanalytic behavior can be determined
without the full machinery of chiral perturbation theory---it depends
only on the existence of dispersion relations and a
pseudo-Goldstone pion.

In principle, an exact evaluation of a current-current correlator in
terms of QCD degrees of freedom must reproduce all of the hadronic
physics, including the physics associated with the existence of a
pseudo-Goldstone pion;
therefore, an ``exact'' implementation of QCD sum rules must be consistent
with chiral perturbation theory.
As noted above, however, all practical implementations are not exact;
thus practical applications of QCD sum rules are not necessarily
consistent with chiral perturbation theory.
By studying the magnitude of this inconsistency, one can estimate the
scale of the uncertainities in QCD sum-rule extractions of spectral
parameters.
We find uncertainties of $\sim 100\,\text{MeV}$ for both the nucleon
mass and $\sigma$ term.
This implies that QCD sum rules, as generally used, are not suitable
for evaluations of the $\sigma$ term.


QCD sum-rule analyses of the nucleon mass \cite{ioffe1}
are based on the time-ordered correlation function $\Pi_N(q)$ defined by
\begin{equation}
\Pi_N(q)\equiv i\int d^4x\,e^{iq\cdot x}
\langle{\rm vac}|T\eta_N(x)\overline{\eta}_N(0)|{\rm vac}\rangle\ ,
\end{equation}
where $|{\rm vac}\rangle$ is the physical nonperturbative vacuum state
and $\eta_N$ is an interpolating field with the quantum numbers of a nucleon.
We take the interpolating field for the proton to be \cite{ioffe1,ioffe2}
\begin{equation}
\eta_p=\epsilon_{abc}(u_a^T C\gamma_\mu u_b)\gamma_5\gamma^\mu d_c\ ,
\label{interp}
\end{equation}
where $u_a$ and $d_a$ are up and down quark fields ($a$ is a color index),
$T$ denotes a transpose in Dirac space,
and $C$ is the charge conjugation matrix.
The interpolating field for the neutron $\eta_n$ is obtained by
interchanging the up and down quark fields.
Lorentz covariance and parity invariance imply that
the Dirac structure of $\Pi_N(q)$ is of the form \cite{bjorken,itzykson}
\begin{equation}
\Pi_N(q)\equiv\Pi_1(q^2)\openone+\Pi_q(q^2)\rlap{/}{q}\ ,
\label{decomp}
\end{equation}
where $\openone$ is the unit matrix in Dirac space,
and $\Pi_1$ and $\Pi_q$ are Lorentz scalar functions of $q^2$ only.

The key ingredients of QCD sum rules are analyticity and asymptotic freedom.
The scalar functions $\Pi_i(s)$ ($i=\{1,q\}$)
are analytic everywhere except on
the positive $s$ axis \cite{bjorken,itzykson}; thus an application of
the Cauchy theorem yields
\begin{equation}
\int_0^R ds\,W_i(s)\Delta\Pi_i(s)
\stackrel{R\rightarrow\infty}{=}-\int_{|s|=R}ds\,W_i(s)\Pi_i^{\rm OPE}(s)\ ,
\label{int1}
\end{equation}
where the weight functions $W_i(s)$ are arbitrary entire functions; the
contour integral on the right-hand side is evaluated in a
counter-clockwise sense.
We take $R$ to be large, so we can evaluate the integral around the
circle on the right-hand side of Eq.~(\ref{int1}) using the OPE \cite{collins}.
The effects of a truncated OPE are considered below.
The discontinuity $\Delta\Pi_i$ is defined by
\begin{equation}
\Delta\Pi_i(s)\equiv\lim_{\epsilon\rightarrow 0^+}
\Pi_i(s+i\epsilon)-\Pi_i(s-i\epsilon)\ .
\end{equation}
Note that $\Delta\Pi_i$ is proportional to the spectral density.

Although the OPE evaluation of $\Pi_i(s)$ is only valid for large $|s|$
\cite{collins},
it is useful to consider the analytic structure of $\Pi_i^{\rm OPE}(s)$
for all $s$.
One finds that $\Pi_i^{\rm OPE}(s)$, like $\Pi_i(s)$, is analytic everywhere
except on the positive $s$ axis; therefore, we can introduce the
following definition:
\begin{equation}
\int_0^R ds\,W_i(s)\Delta\Pi_i^{\rm OPE}(s)
\stackrel{R\rightarrow\infty}{\equiv}
-\int_{|s|=R}ds\,W_i(s)\Pi_i^{\rm OPE}(s)\ .
\label{int2}
\end{equation}
Again, the contour integral on the right-hand side is evaluated in a
counter-clockwise sense.
By combining Eqs.~(\ref{int1}) and (\ref{int2}), we arrive at the
following sum rule:
\begin{equation}
\int_0^\infty ds\,W_i(s)\Delta\Pi_i(s)
=\int_0^\infty ds\,W_i(s)\Delta\Pi_i^{\rm OPE}(s)\ .
\label{sumrule}
\end{equation}
Note that the right-hand side of Eq.~(\ref{sumrule}) is only defined in
the context of Eq.~(\ref{int2}).

In order for the sum rules to be useful,
one must choose the weighting function such that higher terms in the
OPE are suppressed, since a truncated OPE is used in practical calculations.
Such terms ultimately take the form of power corrections, which can be
written schematically as
\begin{equation}
\Pi^{\rm OPE}(s)=\sum_n{C_n\over s^n}\ .
\label{scheme}
\end{equation}
Terms in the OPE proportional to higher-dimensional condensates correspond
to higher values of $n$ in Eq.~(\ref{scheme}).
A straightforward calculation then yields the following form for
the OPE side of the sum rule:
\begin{equation}
\int_0^\infty ds\,W(s)\Delta\Pi^{\rm OPE}(s)
=-2\pi i\sum_n{C_n\over\Gamma(n)}
\left.\left(d\over ds\right)^{n-1}W(s)\right|_{s=0}\ .
\end{equation}
One can use this result to test whether a particular weighting function
efficiently suppresses higher terms in the OPE.
In addition, one must choose the weighting function to suppress the continuum
contributions to the phenomenological side of the sum rule, which can
be accomplished by choosing a $W(s)$ that drops sufficiently rapidly
with increasing $s$.
Of course, there is no guarantee that both goals can be met simultaneously.

The correlator contains contributions from the nucleon pole
and from higher-mass (continuum) states with the quantum numbers of a nucleon.
Lorentz covariance implies that the correlator can be parameterized as
\cite{bjorken,itzykson}
\begin{equation}
\Pi_N(q)=-\lambda_N^2{1\over\rlap{/}{q}-M_N}+\Pi_N^{\rm cont}(q)\ ,
\end{equation}
where $\lambda_N$ specifies the strength of
the coupling between the interpolating field and the physical nucleon
state, and $\Pi_N^{\rm cont}$ denotes the contributions from the continuum.
Thus the scalar functions in the correlator are
\begin{eqnarray}
\Pi_1(s)&=&-\lambda_N^2{M_N\over s-M_N^2}+\Pi_1^{\rm cont}(s)\ ,
\\*
\Pi_q(s)&=&-\lambda_N^2{1\over s-M_N^2}+\Pi_q^{\rm cont}(s)\ .
\end{eqnarray}

On the other hand, one can use the OPE to calculate the scalar
functions in Eq.~(\ref{decomp}).
With the interpolating field in Eq.~(\ref{interp}), one obtains
\cite{ioffe1,ioffe3}
\begin{eqnarray}
\Pi_1^{\rm OPE}(s)&=&
{s\over 4\pi^2}\ln(-s)\langle\overline{q}q\rangle
-{1\over 96\pi^2 s}\langle\overline{q}q\rangle\langle(gG)^2\rangle
+\cdots\ ,
\\*
\Pi_q^{\rm OPE}(s)&=&-{s^2\over 64\pi^4}\ln(-s)
-{1\over 128\pi^4}\ln(-s)\langle(gG)^2\rangle
-{2\over 3s}\langle\overline{q}q\rangle^2+\cdots\ ,
\end{eqnarray}
where $\langle\overline{q}q\rangle\simeq -(225\,\text{MeV})^3$
[$\overline{q}q\equiv{1\over 2}(\overline{u}u+\overline{d}d)$]
and $\langle(gG)^2\rangle\simeq 0.5\,\text{GeV}^4$ are the usual quark and
gluon vacuum condensates \cite{shifman,reinders}.
The ellipses denote contributions from higher-dimensional condensates,
$\alpha_s$ corrections, and contributions proportional
to the current quark mass $m_q\equiv{1\over 2}(m_u+m_d)$.
The latter contributions are not important for our purpose,
since they yield terms that are analytic in $m_q$; we are mainly interested
in terms that are nonanalytic in $m_q$

The usual QCD sum rules based on the Borel transform can be obtained by
choosing $W_1(s)=W_q(s)=W(s)\equiv e^{-s/M^2}$, where $M$ is
known as the Borel mass \cite{shifman,reinders}.
Using the two Borel sum rules for the nucleon, one can readily solve for the
nucleon mass:
\begin{equation}
M_N={\displaystyle{\int_0^\infty dt\,W(t)
\left[\Delta\Pi_1^{\rm OPE}(t)-\Delta\Pi_1^{\rm cont}(t)\right]}
\over
\displaystyle{\int_0^\infty ds\,W(s)
\left[\Delta\Pi_q^{\rm OPE}(s)-\Delta\Pi_q^{\rm cont}(s)\right]}}\ .
\label{mnucsr}
\end{equation}
It is here that one can see a potential limitation of the sum rules.

We consider how the quantities in the sum rules depend on $m_q$ or,
equivalently,
$\overcirc{m}_\pi^2\equiv-2m_q\overcirc{\langle\overline{q}q\rangle}/
\overcirc{f}_\pi^2$.
(All quantities of the form
$\overcirc{x}$ denote the first term in the chiral expansion of $x$.)
The leading terms in the chiral expansion of the nucleon mass are
\begin{equation}
M_N=\overcirc{M}_N+A\overcirc{m}_\pi^2
-{3\overcirc{g}_A^2\over 32\pi\overcirc{f}_\pi^2}\overcirc{m}_\pi^3
+\cdots\ ,
\label{mnucexp}
\end{equation}
where $A$ is an unknown constant \cite{gasser2}.
The expansion in Eq.~(\ref{mnucexp}) includes the
leading nonanalytic term in $m_q$.
In principle, the QCD sum-rule result for $M_N$ [Eq.~(\ref{mnucsr})]
reflects this expansion.

In practice, however, it is difficult to obtain sum-rule results for
the nucleon mass that are consistent with Eq.~(\ref{mnucexp}).
According to Eq.~(\ref{mnucsr}), the $O(\overcirc{m}_\pi^3)$
contribution to the nucleon mass, which is nonanalytic in the current
quark mass, must come from either the OPE or the continuum.
We expect that it must be from the continuum; standard OPE calculations
based on perturbation theory yield a sum of Wilson coefficients that
are analytic in the current quark mass multiplied by vacuum
condensates.
We know of no vacuum condensate with an $O(\overcirc{m}_\pi^3)$
contribution.

There are more serious difficulties.
The leading chiral behavior of the quark condensate is \cite{novikov}
\begin{equation}
\langle\overline{q}q\rangle=\overcirc{\langle\overline{q}q\rangle}
\left(1-{3\over 32\pi^2\overcirc{f}_\pi^2}\overcirc{m}_\pi^2
\ln{\overcirc{m}_\pi^2\over M_0^2}+\cdots\right)\ ,
\label{qbarqexp}
\end{equation}
where $M_0$ is a mass parameter that is presumably $\sim 1\,\text{GeV}$.
We take $M_0$ to be in the range 500--1500 MeV.
The quark condensate given in Eq.~(\ref{qbarqexp}) is one in which
perturbative contributions have been subtracted off \cite{novikov};
thus it is presumably suitable for use in the OPE.

Thus $\Delta\Pi_1^{\rm OPE}$ contains an
$O(\overcirc{m}_\pi^2\ln\overcirc{m}_\pi^2)$ chiral logarithm that must
be eliminated somewhere in the sum-rule expression for $M_N$
[Eq.~(\ref{mnucsr})], since it does not appear in the chiral
perturbation theory expression for $M_N$ [Eq.~(\ref{mnucexp})].
It is clear that the contributions to the OPE correlator from
higher-dimensional condensates will not eliminate this chiral
logarithm, since they are multiplied by different powers of $1/q^2$;
therefore, the chiral logarithm must be canceled by contributions to
the continuum.

Thus we expect the continuum to play an important role in guaranteeing the
correct chiral behavior of the nucleon mass as deduced by QCD sum rules.
Standard treatments of QCD sum rules, however, assume a very simple form
for the continuum \cite{shifman,reinders}:
\begin{equation}
\Delta\Pi_i^{\rm cont}(s)=\theta(s-s_0^i)\Delta\Pi_i^{\rm OPE}(s)\ ,
\end{equation}
where $s_0^i$ are continuum threshold parameters.
We take $s_0^1=s_0^q=s_0$.
With this ansatz, Eq.~(\ref{mnucsr}) becomes
\begin{equation}
M_N={\displaystyle{\int_0^{s_0}dt\,W(t)\Delta\Pi_1^{\rm OPE}(t)}
\over
\displaystyle{\int_0^{s_0}ds\,W(s)\Delta\Pi_q^{\rm OPE}(s)}}\ .
\label{mnucsimple}
\end{equation}
Such a simple treatment of the continuum will almost
invariably lead to a spurious chiral logarithm in the resulting nucleon
mass due to the $O(\overcirc{m}_\pi^2\ln\overcirc{m}_\pi^2)$ chiral
logarithm in $\langle\overline{q}q\rangle$.
In addition, the needed $O(\overcirc{m}_\pi^3)$ contribution
to the nucleon mass will almost certainly be lost.
It is clear that one must describe the continuum contribution to the
correlator quite exactly in order to obtain chiral behavior that is
consistent with Eq.~(\ref{mnucexp}).

What are the consequences of the spurious and missing nonanalytic terms
in QCD sum-rule calculations of the nucleon mass?
This question can be examined by an explicit Borel sum-rule
calculation.
Solving Eq.(\ref{mnucsimple}) leads to the the following form:
\begin{equation}
M_N=g(s_0;M^2)\langle\overline{q}q\rangle+\cdots\ ,
\label{ioffe}
\end{equation}
where the ellipsis denotes contributions from higher-dimensional condensates.
We have shown that the leading term in $\Pi_1^{\rm OPE}$, which is
proportional to $\langle\overline{q}q\rangle$, leads to a spurious
chiral logarithm contribution to the nucleon mass, given the usual
simple ansatz for $\Pi_1^{\rm cont}$.
Contributions to the OPE correlator from higher-dimensional condensates do not
resolve the problem, since they are multiplied by different powers of $1/q^2$;
therefore, for our purposes, we can neglect these contributions.
Clearly Eq.~(\ref{ioffe}) is not adequate for a precise determination of the
nucleon mass; however, it should be sufficient for determining the scale
of the uncertainty in sum-rule estimates of $M_N$ due to spurious and
missing nonanalytic terms.

The nucleon mass, as determined by chiral perturbation theory, consists of
unknown analytic (in $m_q$) and known nonanalytic (in $m_q$) contributions.
In particular, from Eq.~(\ref{mnucexp}), one has
\begin{equation}
M_N^{\rm XPT}=\text{analytic contributions}
-{3\overcirc{g}_A^2\over 32\pi\overcirc{f}_\pi^2}\overcirc{m}_\pi^3+\cdots\ .
\label{xptnonanal}
\end{equation}
It is believed that Eq.~(\ref{xptnonanal}) gives the correct expansion of
the nucleon mass \cite{gasser2}.
The QCD sum-rule estimate of the nucleon mass, as given in
Eq.~(\ref{ioffe}), can also be written in terms of analytic and
nonanalytic contributions:
\begin{equation}
M_N^{\rm QCDSR}=\text{analytic contributions}
-{3\overcirc{M}_N\over 32\pi^2\overcirc{f}_\pi^2}
\overcirc{m}_\pi^2\ln{\overcirc{m}_\pi^2\over M_0^2}+\cdots\ ,
\label{qcdsrnonanal}
\end{equation}
where we have used Eq.~(\ref{qbarqexp}) and Eq.~(\ref{ioffe}) in the
chiral limit.
While we recognize that the numerical value of the continuum
threshold $s_0$ will vary with different choices of $m_q$, we know of
no realistic way to determine a chiral expansion of $s_0$ analogous to
the chiral expansions of $M_N$ and $\langle\overline{q}q\rangle$ given
in Eqs.~(\ref{mnucexp}) and (\ref{qbarqexp}), respectively.
It is difficult to describe the $m_q$ dependence of $s_0$, since it is
an effective parameter in a simple continuum ansatz.
Therefore, in deriving Eq.~(\ref{qcdsrnonanal}), we have neglected the
$m_q$ dependence of $s_0$.
We will return to this issue later.

It is possible that numerical differences in the analytic contributions
to Eqs.~(\ref{xptnonanal}) and (\ref{qcdsrnonanal}) will partially
compensate for the differences in the nonanalytic contributions to
these equations; however, this would be purely fortuitous.
We must assume that the differences in the nonanalytic terms lead to an
intrinsic {\it uncertainty\/} $\delta M_N$ in the QCD sum-rule estimate
of the nucleon mass:
\begin{equation}
\delta M_N\simeq-{3\overcirc{M}_N\over 32\pi^2\overcirc{f}_\pi^2}
\overcirc{m}_\pi^2\ln{\overcirc{m}_\pi^2\over M_0^2}
+{3\overcirc{g}_A^2\over 32\pi\overcirc{f}_\pi^2}\overcirc{m}_\pi^3
\simeq\text{65--110 MeV}\ ,
\end{equation}
where, in obtaining the numerical estimate, we have used physical values
for all quantities.
Thus the uncertainty is fairly modest if one is simply trying to
determine a rough numerical value for the nucleon mass.


Unfortunately, this is {\it not\/} the case with the nucleon $\sigma$ term,
which can be defined as \cite{gasser3}
\begin{equation}
\sigma_N\equiv2m_q\int d^3 x\,(\langle N|\overline{q}q|N\rangle
-\langle{\rm vac}|\overline{q}q|{\rm vac}\rangle)\ .
\end{equation}
The state vector $|N\rangle$ is for a nucleon at rest.
The most recent value of the $\sigma$ term is
$45\pm 10\,\text{MeV}$ \cite{gasser3}.

By an application of the Hellmann-Feynman theorem, one can show \cite{cohen}
\begin{equation}
\sigma_N=m_q{dM_N\over dm_q}
=\overcirc{m}_\pi^2{d\overcirc{M}_N\over d\overcirc{m}_\pi^2}\ .
\label{hf}
\end{equation}
The latter form is most useful for our purposes.
Evaluating the derivative in Eq.~(\ref{hf}) using a QCD sum-rule
estimate of the nucleon mass leads to a QCD sum-rule
estimate of the nucleon $\sigma$ term.
This approach is equivalent to that used in Ref.~\cite{jin}, which is based on
evaluating the nucleon correlator in the presence of a background scalar
field.

In analogy to Eq.~(\ref{xptnonanal}), we can write the chiral expansion of the
$\sigma$ term as
\begin{equation}
\sigma_N^{\rm XPT}=\text{analytic contributions}
-{9\overcirc{g}_A^2\over 64\pi\overcirc{f}_\pi^2}\overcirc{m}_\pi^3+\cdots\ .
\end{equation}
Likewise, from Eq.~(\ref{qcdsrnonanal}), one obtains the following expansion
of the QCD sum-rule estimate of the $\sigma$ term:
\begin{equation}
\sigma_N^{\rm QCDSR}=\text{analytic contributions}
-{3\overcirc{M}_N\over 32\pi^2\overcirc{f}_\pi^2}
\overcirc{m}_\pi^2\ln{\overcirc{m}_\pi^2\over M_0^2}+\cdots\ .
\end{equation}
Thus QCD sum-rule estimates of the nucleon $\sigma$ term that are based on
a simple ansatz for the continuum have an uncertainty
$\delta\sigma_N$ given by
\begin{equation}
\delta\sigma_N\simeq-{3\overcirc{M}_N\over 32\pi^2\overcirc{f}_\pi^2}
\overcirc{m}_\pi^2\ln{\overcirc{m}_\pi^2\over M_0^2}
+{9\overcirc{g}_A^2\over 64\pi\overcirc{f}_\pi^2}\overcirc{m}_\pi^3
\simeq\text{70--115 MeV}\ .
\end{equation}
Unlike the case of the nucleon mass, the QCD sum-rule determination of
the nucleon $\sigma$ term is completely unsatisfactory.
The uncertainty in the QCD sum-rule estimate of the $\sigma$ term
associated with chiral physics is about two times the value of the
$\sigma$ term itself.

At this stage, it is worth returning to the issue of the $m_q$
dependence of $s_0$.
One might hope that a nontrivial dependence exists and that such a
dependence can eliminate or reduce the mismatch between chiral
perturbation theory and the QCD sum-rule description.
For a given choice of weighting function, one can, in fact, fabricate
an $m_q$ dependence of $s_0$ that {\it does\/} eliminate the
discrepancy.
However, such a solution is completely arbitrary; we have no
fundamental reason to believe that this occurs.
Moreover, this solution to the problem is unsatisfactory, since the
$m_q$ dependence that eliminates the mismatch depends on the choice of
weighting function.
Furthermore, in actual implementations of QCD sum rules, $s_0$ is
determined via some optimization procedure.
Such a procedure is extremely unlikely to give rise to the $m_q$
dependence needed to achieve consistency with the correct chiral
behavior.

In conclusion, we have shown intrinsic uncertainities of
$\sim 100\,\text{MeV}$ in QCD sum-rule extractions of the nucleon mass and
$\sigma$ term associated with inconsistencies between the sum rules and
known chiral behavior.
As a practical matter, this suggests that attempts to improve QCD
sum-rule calculations for the nucleon via an improved treatment of the
OPE or a more judicious choice of weighting function are of limited value in
reducing the uncertainties.
We see from the present study that a 100-MeV uncertainty
will persist until the inadequacies of the continuum ansatz are
addressed.


\acknowledgements

The authors thank M. Banerjee, H. Forkel, R. Furnstahl, and M. Nielsen
for useful conversations.
Support from the Department of Energy under
Grant No.\ DE--FG02--93ER--40762 is gratefully acknowledged.
T.D.C. acknowledges additional support from the National Science
Foundation under Grant No.\ PHY--9058487.



\end{document}